\documentclass[letter]{elsart}
\usepackage{amsmath}
\usepackage{latexsym}
\usepackage{amssymb}
\usepackage{type1cm}

\def\Tr{\mathop{\rm Tr}\nolimits}

\def\SU{\mathop{\rm SU}\nolimits}

\newcommand{\defeq}{\stackrel{\rm def}{=}}

\def\cH{\mathcal{H}}

\def\cD{\mathcal{D}}
\def\Label#1{\label{#1}\ [\ #1\ ]\ }
\def\Label{\label}
\def\cho#1#2{{\displaystyle #1 \atopwithdelims() \displaystyle #2}}
\journal{Physics Letters A}
\begin{document}
\begin{frontmatter}
\title{Parallel Treatment of\\ Estimation of SU(2) and Phase Estimation\thanksref{label1}}
\author{Masahito Hayashi}
\ead{masahito@qci.jst.go.jp}
\address{Quantum Computation and Information Project, ERATO, JST\\
5-28-3, Hongo, Bunkyo-ku, Tokyo, 113-0033, Japan}
\thanks[label1]{The material in 
this paper was presented in 
part at The Ninth Quantum Information Technology Symposium (QIT9),
Atsugi, Kanagawa, Japan, December 11--12, 2003.
It was also presented in 
part at
The Seventh International Conference on
Quantum Communication, Measurement and Computing,
Glasgow, UK, July 25--29, 2004.
The final section (Concluding remark) was added after 
these presentations.}
\begin{abstract}
We discuss the accuracy of the estimation of the $n$ identical
unknown actions of $\SU(2)$
by using entanglement.
This problem has a similar structure 
with the phase estimation problem, which was discussed by Bu\v{z}ek, 
Derka, and Massar\cite{BDM}.
The estimation error asymptotically decreases to zero
with an order of $\frac{1}{n^2}$ at least.
\end{abstract}

\end{frontmatter}
\maketitle

\section{Introduction}
\subsection{Estimation of unitary action}
Quantum information processing architecture
is constructed with a combination of various quantum devices.
In other words, each quantum device contributes towards
obtaining the output quantum state from the input quantum state.
Hence, it is necessary to identify 
the quantum operation acting on each device.
This problem is referred to as the identification of the quantum operation;
this problem usually entails the estimation 
of the operation based on the pair of the input (initial) state 
and the data of the measurement for the output state.
Generally, such a quantum operation is described by a
trace-preserving completely positive (TP-CP) map.
In this paper, we consider a noiseless case
in which no noise appears in the quantum operation.
In this case, the quantum operation is described by a unitary matrix.
The adiabatic case is also treated in a similar manner.

When the reference system is unitarily equivalent to the input system 
and the initial state is 
a maximally entangled state in a joint system
between the original input system and
the reference system,
the final state in the joint system
is the maximally entangled state
described by the unitary matrix acting on the input system.
Hence, by repeating this operation $n$ times,
we can estimate this unitary matrix by performing appropriate measurement
for the total joint system between output system and
the reference system.
In the two-dimensional case, Fujiwara \cite{Fuji}
showed that a maximally entangled state
is the optimal initial state in the composite system between 
the single input system and its reference system.
Ballester \cite{Bal} discussed this problem in the $d$-dimensional case.
When their method is applied to the case of $n$ applications 
of this unknown operation,
the estimation error decreases to zero only with 
order $\frac{1}{n}$;
this is the case even if the output measurement is optimized.
This order is obtained from the accuracy 
of state estimation by using an identical state preparation.
Therefore, it is worthwhile to consider 
the possibility of further improvement.

\subsection{One-parameter case: Phase estimation}
The phase estimation problem is closely related to the problem
discussed in this paper.
In the phase estimation problem,
we estimate the eigenvalue $e^{i\theta}$ 
of the unknown unitary matrix
acting on a two-level system 
when we know its eigenvectors.
The phase estimation with the fixed input state 
was discussed by Helstrom\cite{Hel} first.
Holevo \cite{H1} extended Helstrom's result to 
a more general framework of group covariance.
The optimization problem of the input state for phase estimation 
was discussed by Bu\v{z}ek {\it et al.} \cite{BDM} in an asymptotic setting.
They proved that 
the error decreases to zero with speed $\frac{\pi^2}{4 n^2}$ 
when we choose the optimal input state and optimal measurement.
However, the error goes to zero with the order $\frac{1}{n}$
in the usual statistical parameter estimation.
Therefore, this unexpected result 
indicates the importance of the entangled input state.

\subsection{Three-parameter case: Our result}
In this study, we discuss 
whether such a phenomenon occurs in the estimation of SU(2) unitary action.
In this paper, 
we adopt the error function 
$d(U,\hat{U})\defeq 1- |\Tr \frac{U^{-1}\hat{U}}{2}|^2$
between the true SU(2) matrix $U$ and the estimated matrix $\hat{U}$.
Then, we obtain an unexpected relation
between our problem and that discussed by Bu\v{z}ek {\it et al.}\cite{BDM}.
Thanks to this relation,
we can show that
the error decreases to zero in proportion to $\frac{1}{n^2}$ at least.
Further, the coefficient is equal to $\pi^2$
if the estimator is constructed based on this relation.
It is shown that
this bound can be asymptotically attained with no use of the reference system.
Instead of use of the reference system,
we regard a part of the composite system of 
$n$ input systems as the tensor product of 
the system of interest and the reference system.
In other words, 
an effect of ``self-entanglement'' is used in this method.

\section{Phase estimation: Estimation of eigenvalues with the knowledge of eigenvectors}
In order to demonstrate the interesting relation between
phase estimation and estimation of $\SU(2)$ action,
we briefly summarize the fact known with regard to phase estimation.
When the eigenvectors of the unknown $\SU(2)$ matrix
$U$ are known,
the estimation problem can be reduced to
the estimation of 
the unknown parameter $\theta\in [0,2\pi)$ of the family
$\left\{\left. U_\theta \defeq
\left(
\begin{array}{cc}
e^{i\theta} & 0 \\
0 & 1
\end{array}
\right)
\right| \theta \in [0,2 \pi)\right\}
$ because the action of $U$ is equivalent to that of $cU$.

In our framework, we assume that 
the tensor product matrix $U_\theta^{\otimes n}$ acts in the
tensor product space $\cH^{\otimes n}$ and 
that we can select an arbitrary initial state in the 
tensor product space $\cH^{\otimes n}$.
That is,
we can select the input state $x \in \cH^{\otimes n}$ and
the estimating POVM $M^n(d \hat{\theta})$ on $\cH^{\otimes n}$
with outcomes in $[0,2\pi)$.
When we evaluate the error between the true parameter $\theta$ and
the estimated parameter $\hat{\theta}$ by
the function $\sin^2\frac{\theta-\hat{\theta}}{2}$,
the mean error is given by
\begin{eqnarray*}
 {\cD}_\theta^n(M^n, x)
\defeq 
\int_0^{2\pi}
\sin^2\frac{\theta-\hat{\theta}}{2}
\langle x|
(U_\theta^{\otimes n})^\dagger M^n(d \hat{\theta}) 
U_\theta^{\otimes n} |x \rangle .
\end{eqnarray*}
Thus, when we fix the input state $x\in \cH^{\otimes n}$,
our estimation problem can be reduced to
the estimation with the state family 
$
\left\{\left.U_\theta^{\otimes n} |x \rangle \langle x|
(U_\theta^{\otimes n})^\dagger
\right|
\theta \in [0,2\pi)\right\}$.

In order to treat this problem in a simple manner,
we focus on the phase estimation problem in 
the $d$-dimensional space
$\cH'$ spanned by the
orthonormal basis $u_1,\ldots, u_d$ as follows.
We select a vector $x=(x_k)_{k=1}^d=\sum_{k=1}^d x^k u_k$ satisfying
\begin{eqnarray}
\sum_{k=1}^{d} |x_k|^2=1 .\Label{3-12-5}
\end{eqnarray}
Suppose that the state to be estimated 
has the form $\rho_{\theta,x} 
\defeq U_\theta |x \rangle\langle x|U_\theta^*$
with the unknown parameter 
$\theta\in [0,2\pi)$ and
the unitary matrix $U_\theta\defeq \sum_{k=1}^d e^{i k \theta}| u_k \rangle
\langle u_k|$.
In this case, 
the error of the estimator $M(\,d \hat\theta)$
can be be expressed as
\begin{eqnarray}
{\cD}_\theta(M,x)\defeq
\int_0^{2\pi} \sin^2\frac{\theta-\hat{\theta}}{2}
\Tr M(d \hat{\theta})\rho_{\theta,x}.
\end{eqnarray}
In order to ensure accuracy, 
it is appropriate to focus on the worst case.
In other words, we minimize 
the worst error $\max_{\theta}{\cD}_\theta(M,x)$.
This problem is refereed to as minimax.

On the other hand,
since the state $\rho_{\theta,x}$ has symmetry
\begin{eqnarray}
\rho_{\theta+\theta',x} = U_{\theta'}\rho_{\theta,x} 
U_{\theta'}^\dagger ,
\Label{10-20}
\end{eqnarray}
it is natural to treat the measurement 
$M(\hat{\theta}) d \hat{\theta}$
with the same symmetry:
\begin{eqnarray}
M(\hat{\theta} +\hat{\theta}') 
= U_{\hat{\theta}'} M(\hat{\theta}) U_{\hat{\theta}'}^\dagger .
\label{11-22-1-q}
\end{eqnarray}
A measurement $M$ satisfying (\ref{11-22-1-q})
is referred to as the covariant measurement and 
it has the form
\begin{eqnarray}
M^T(d \hat{\theta})\defeq
U_{\hat{\theta}} T U_{\hat{\theta}}^\dagger 
\frac{d \hat{\theta}}{2 \pi}, \Label{3-12-7}
\end{eqnarray}
where the Hermitian matrix 
$T=\sum_{k,l} t_{k,l}|u_{k}\rangle \langle u_{l}|$
satisfies
\begin{eqnarray}
T \ge 0 \hbox{ and }
t_{k,k}= 1. \Label{3-11-2}
\end{eqnarray}

Holevo proved that the minimum error in the minimax criteria
is equal to the minimum error among covariant measurements,
{\it i.e.},
\begin{eqnarray*}
\min_M \max_{\theta}{\cD}_\theta(M,x)
=
\min_{M:{\rm covariant}} {\cD}_\theta(M,x).
\end{eqnarray*}
This relation is referred as the quantum Hunt-Stein lemma, and was
proved in a more general covariant setting\cite{H1}.

By using elementary formulas of trigonometric functions,
the equation
\begin{eqnarray}
&\int_0^{2\pi} \sin^2\frac{\theta-\hat{\theta}}{2}
\Tr 
U_\theta |u_{k'} \rangle \langle u_{l'}|U_\theta^\dagger 
U_{\hat{\theta}} |u_l \rangle \langle u_k| U_{\hat{\theta}}^\dagger
\frac{d \hat{\theta}}{2 \pi}\nonumber \\
=& \delta_{k,k'}\delta_{l,l'}
(\frac{1}{2}\delta_{k,l}- \frac{1}{4}\delta_{k,l-1}
- \frac{1}{4}\delta_{k-1,l}) \Label{3-12-8}
\end{eqnarray}
can be verified.
By using this relation,
we have
\begin{align}
&  {\cD}_\theta(M_T,x)
=\frac{1}{2}\sum_{k=1}^d |x_k|^2 t_{k,k}
- \frac{1}{4}\sum_{k=1}^{d-1} 
\left(\overline{x_{k}} x_{k+1} t_{k+1,k}
+\overline{x_{k+1}} x_{k} t_{k,k+1} \right)
\nonumber\\
 \ge &
\frac{1}{2}\sum_{k=1}^d |x_k|^2 
- \frac{1}{4}\sum_{k=1}^{d-1} 
\left(|\overline{x_{k}} x_{k+1} |
+|\overline{x_{k+1}} x_{k}|\right)
=  \frac{1}{2}\left(
1- \sum_{k=1}^{d-1} |x_k||x_{k+1}|
\right), \Label{3-12-9}
\end{align}
where the inequality follows from 
$|\langle u_{k}| T |u_{k+1} \rangle|\le 
\sqrt{\langle u_k| T |u_k \rangle
\langle u_{k+1}| T |u_{k+1} \rangle}=1$.
Hence, the equality holds iff 
\begin{eqnarray}
t_{k,k+1} 
= \frac{\overline{x_{k}} x_{k+1}}{|x_k| |x_{k+1}|}. \Label{3-11-1}
\end{eqnarray}
Since the matrix
$T_x \defeq \sum_{k,l}\frac{\overline{x_k} x_l}{|x_k| |x_l|}
|u_{k} \rangle\langle u_l|$
satisfies the conditions (\ref{3-11-2}) and (\ref{3-11-1}),
we obtain
\begin{eqnarray}
\min_{M:{\rm covariant}} {\cD}_\theta(M,x)
=
\frac{1}{2}\left(
1- \sum_{k=1}^{d-1} |x_k||x_{k+1}|
\right).\Label{3-11-3}
\end{eqnarray}

Now, we consider the estimation of the unknown unitary $U_\theta$
with its multiple actions.
Since the unitary matrix $U_\theta^{\otimes n}$ has
eigenvalues $1, e^{i \theta} , \ldots, e^{n i \theta}$,
the application of the relation (\ref{3-11-3}) to the $d=n+1$ case
yields 
\begin{align*}
{\cD}_{opt}^n 
\defeq  &\min_{x\in\cH^{\otimes n}}\min_{M^n}
\max_\theta{\cD}_\theta^n(M^n, x)\\
= &\min_{(a_k)_{k=0}^n }
\left\{\left.
\frac{1}{2}
\left( 1- \sum_{k=1}^n a_k a_{k-1}\right)
\right|
\sum_{k=0}^n a_k^2 =1, a_k \ge 0
\right\}.
\end{align*}
In this problem,
the eigenspace $\cH_k$ is not one-dimensional for the eigenvalue
$e^{i k \theta}$.
Hence, any initial state can be written as 
$\sum_k x_k e^k$, where $e^k$ is a normalized vector in $\cH_k$.
Therefore, the minimum error can be expressed in the above form.
Bu\v{z}ek, {\it et al}.\cite{BDM} proved that
the minimum error is attained by the coefficient 
$a_k= \frac{\sqrt{2}}{\sqrt{n+1}}$ $\sin\frac{\pi (k+1/2)}{n+1}$
and is almost equal to $\frac{\pi^2}{4 n^2}$,
{\it i.e.},
\begin{eqnarray}
{\cD}_{opt}^n 
\cong \frac{\pi^2}{4 n^2}.\Label{3-12-1}
\end{eqnarray}

In this setting, 
the irreducible representation space of 
the action of the one-dimensional circle $S^1=[0,2\pi)$
is the one-dimensional space spanned by $u_k$.
In the above discussion, 
we use $n+1$ different irreducible representation spaces.

\section{Estimation of the unknown $\SU(2)$ action}
Next, we proceed to the estimation of the unknown $\SU(2)$ action 
$g \in \SU(2)$ in the two-dimensional space $\cH$.
In this problem, the unitary matrix $g^{\otimes n}$ 
acts in the tensor product space $\cH^{\otimes n}$,
and a suitable initial state in $\cH^{\otimes n}$
can be selected for this estimation.
On the other hand,
Fujiwara\cite{Fuji} focused on 
the estimation problem where the initial state is 
entangled with the reference system $\cH_R$,
in which the unknown $\SU(2)$ action $g$ does not act.
He proved that this method reduces the estimation error.
However, he did not treat the entanglement 
between the $n$-tensor product space $\cH^{\otimes n}$
and its reference system.

Now, we consider the amount of the estimation error
can be reduced by use of 
entanglement among tensor product space.
In other words, our framework has a wider choice.
In our problem, we can select an initial state 
entangled between $\cH^{\otimes n}$ and the 
reference space $\cH_{R}^{\otimes n}$ 
that is unitarily equivalent to the original input system $\cH^{\otimes n}$.
Hence, when we select the initial state $x$ on the tensor product space
$\cH^{\otimes n}\otimes \cH_{R}^{\otimes n}$ 
and the measurement (POVM) $M^n(d \hat{g})$ in 
$\cH^{\otimes n}\otimes \cH_{R}^{\otimes n}$ with the outcome in $\SU(2)$,
the error is evaluated as
\begin{eqnarray*}
{\cD}_g(M,x)
\defeq 
\!\int_{\SU(2)}\! d(g,\hat{g})
\langle x| (g^{\otimes n}\otimes I)^*
M(d \hat{g})(g^{\otimes n}\otimes I)|x \rangle .
\end{eqnarray*}

Here, we focus on the $\SU(2)$ action on the tensor product space
$\cH^{\otimes n}$.
Its irreducible decomposition is given as follows:
\begin{align*}
\cH^{\otimes 2d}
& =
\bigoplus_{k=0}^d \cH_{2k+1} \otimes \cH_{2d,2k+1} \\
\cH^{\otimes 2d-1}
& =
\bigoplus_{k=1}^d \cH_{2k} \otimes \cH_{2d-1,2k} ,
\end{align*}
where $\cH_k$ is the $k$-dimensional irreducible space of the 
$\SU(2)$ action,
and $\cH_{n,k}$ is the corresponding irreducible space of
the action of the permutation group,
where $\SU(2)$ does not act.
Note that the dimension of $\cH_{n,k}$ is equal to 
the number of representation spaces equivalent to $\cH_k$
in the tensor product space $\cH^{\otimes n}$.
Hereafter, we denote the $\SU(2)$ action on $\cH_k$ by $V_{g}^k$.

For simplicity, first we focus on 
the estimation of the $\SU(2)$ action in a single irreducible
space $\cH_{j}$.
Now, we select the initial state as the maximally entangled state
$x^j_{E}$ between $\cH_j$ and the reference space $\cH_{j,R}$
that is unitarily equivalent with the space $\cH_j$.
The measurement is selected as 
the POVM $M_{j,E}(d \hat{g})\defeq j^2
V_{\hat{g}}^j \otimes I|x^j_{E} \rangle\langle x^j_{E}| 
(V_{\hat{g}}^j\otimes I)^\dagger
\mu(d \hat{g})$,
where $\mu$ is the invariant probability distribution on $\SU(2)$
and the integer $j^2$ is the normalizing factor.
By using Schur's lemma, 
we can easily verify that
the total integral is constant times the identity matrix
because the state $x^j_{E}$ is maximally entangled.
Since $(\Tr V_{g}^j)^* = \Tr (V_{g}^j)^\dagger$,
the average error is calculated as
\begin{align}
& \int_{\SU(2)} d(g,\hat{g})
j^2
\langle x^j_{E}| (V_g^j)^\dagger 
V_{\hat{g}}^j |x^j_{E} \rangle \langle x^j_{E}| (V_{\hat{g}}^j)^\dagger 
V_g^j 
|x^j_{E} \rangle \mu(d \hat{g}) \nonumber\\
= &
\int_{\SU(2)} d(g,\hat{g})
|\Tr V_{\hat{g}^{-1} g}^j|^2 \mu(d \hat{g}) 
= 
\int_{\SU(2)} d(I,\hat{g})
|\Tr V_{\hat{g}^{-1}}^j|^2 \mu(d \hat{g}) \nonumber\\
= &
\int_{\SU(2)} d(I,\hat{g})
|\Tr V_{\hat{g}}^j|^2 \mu(d \hat{g}) 
=\left\{
\begin{array}{cc}
\frac{3}{4} & j= 1 \\
\frac{1}{2} & j \ge 2 .
\end{array}
\right. 
\Label{2}
\end{align}
The final equation follows from the elementary calculus of
trigonometric functions.
This calculation seems to indicate that 
if we use only one irreducible space $\cH_{j}$
and even if its dimension is large,
the estimation error cannot be reduced\footnote{This 
fact has been proved by Chiribella {\it et al}.\cite{CDF}
after the submission of the preliminary version
of this paper. That is, they proved the optimality of this POVM
in a more general framework.}.
Hence, in order to reduce the estimation error,
it may be needed to use several irreducible spaces.

Hereafter, we consider the case of $n=2d-1$;
however, the following discussion can be applied to the even case.
In the odd case, $d$ distinct irreducible spaces exist.
Hence, it is essential to use the correlation between them.
We investigate the following subspace of 
$\cH^{\otimes (2d-1)}\otimes\cH_{R}^{\otimes (2d-1)}$:
\begin{eqnarray*}
\bigoplus_{k=1}^m \cH_{2k} \otimes \cH_{2k,R}.
\end{eqnarray*} 
Further, we denote the $\SU(2)$ representation on this space 
by $U_g^{2d-1}$.

As demonstrated later,
this problem can be treated parallel to the phase estimation
in which the base $u_j$ corresponds to 
the maximally entangled state $x^{2k}_{E}$
or the vector $(2k) x^{2k}_{E}$.
We select a vector $\vec{x}_d=(x_k)_{k=1}^d$ 
that satisfies the
condition (\ref{3-12-5}),
and let the initial state be
$x_{\vec{x}_d }^{2d-1}\defeq \bigoplus_{k=1}^d x_k x^{2k}_{E}$.
In a manner similar to (\ref{3-12-7}),  the measurement
is selected as
\begin{eqnarray*}
M_{2d-1}^{T}(d \hat{g}) &\defeq
U_{\hat{g}}^{2d-1} 
\sum_{k,l}t_{k,l}
|(2k) x^{2k}_{E} \rangle \langle (2l) x^{2l}_{E}| 
(U_{\hat{g}}^{2d-1} )^\dagger
\mu(d \hat{g})
\end{eqnarray*}
based on a Hermitian matrix $T=(t_{k,j})$ satisfying condition 
(\ref{3-11-2}).
Using Schur's lemma and condition (\ref{3-11-2}),
we can verify that its total integral is an identity matrix.
In a manner similar to (\ref{3-12-8}), the equations
\begin{align}
& \int_{\SU(2)}
d(g,\hat{g})
\Tr
U_g^{2d-1}
|x^{2k'}_{E} \rangle
\langle x^{2l'}_{E}|  
(U_{g}^{2d-1})^\dagger
U_{\hat{g}}^{2d-1} 
|(2k) x^{2k}_{E} \rangle \langle (2l) x^{2l}_{E}| 
(U_{\hat{g}}^{2m-1} )^\dagger
\mu(d \hat{g})\nonumber \\
 =&
\delta_{k,k'}\delta_{l,l'}
\int_{\SU(2)}
d(g,\hat{g})
\Tr V_{\hat{g}}^{2k}
(\Tr V_{\hat{g}}^{2l})^\dagger
\mu(d \hat{g}) 
=
\delta_{k,k'}\delta_{l,l'}
\left(
\frac{1}{2}\delta_{k,l}
-\frac{1}{4}\delta_{k,l-1}-\frac{1}{4}\delta_{k-1,l}
\right)\Label{3}
\end{align}
holds, where the final equation is derived in section \ref{s4} based 
on elementary formulas of trigonometric functions.
By using this relation in a manner similar to (\ref{3-12-9}), we have
\begin{align}
& \int_{\SU(2)}
d(g,\hat{g})
\Tr
U_g^{2d-1}
|x_{\vec{x}_d }^{2d-1}\rangle 
\langle x_{\vec{x}_d }^{2d-1}|(U_{g}^{2d-1})^\dagger
M_{2d-1}^{T}(d \hat{g})\nonumber \\
 =& \frac{1}{2}\sum_{k=1}^d |x_k|^2 t_{k,k}
- \frac{1}{4}\sum_{k=1}^d 
\left(\overline{x_{k-1}} x_k t_{k,k-1}
+\overline{x_{k}} x_{k-1} t_{k-1,k}\right)
\nonumber \\
 \ge &
\frac{1}{2}\sum_{k=1}^d |x_k|^2 
- \frac{1}{4}\sum_{k=1}^{d-1} 
\left(|\overline{x_{k}} x_{k+1} |
+|\overline{x_{k+1}} x_{k}|\right)
=  \frac{1}{2}\left(
1- \sum_{k=1}^{d-1} |x_k||x_{k+1}|
\right).\Label{3-15}
\end{align}
The equality holds when matrix $T=(t_{k,l})$ satisfies 
(\ref{3-11-1}).
Thus, the optimal error of this estimation method 
coincides with ${\cD}_{opt}^{m-1}$.
That is, our problem is reduced to  
phase estimation problem by Bu\v{z}ek {\it et al}.
When we select a suitable 
initial state and measurement in the case of large $n$,
the estimation error 
is equal to $\frac{\pi^2}{n^2}$ asymptotically 
because ${\cD}_{opt}^{d-1}
\cong \frac{\pi^2}{4(d-1)^2}
\cong \frac{\pi^2}{(2d-1)^2}=\frac{\pi^2}{n^2}$.

In the case $n= 2d$,
the initial state is expressed as 
$x_{\vec{a}_d' }^{2d}\defeq \bigoplus_{k=0}^d a_k x^{2k+1}_{E}$,
where the vector $\vec{a}_d'=(a_k)_{k=0}^d$ has no negative element.
When we select a measurement similar to the one mentioned above,
the estimation error is calculated to be
$\frac{1}{2}\left(1- \sum_{k=0}^{d-1} 
a_k a_{k+1}\right)+\frac{1}{4}a_0$.
Hence, the minimum error ${{\cD}_{opt}^{d}}'$ satisfies 
${\cD}_{opt}^{d}\le {{\cD}_{opt}^{d}}'\le {\cD}_{opt}^{d-1}$.
The same conclusion is obtained in the odd case.

Next, we investigate the reference space in the odd-dimensional case.
When we use the above method,
the dimension of the reference space $\cH_{R,2d-1}$ is
$2d$.
If the dimension of $\cH_{2d-1,2k}$ is greater than
that of $\cH_{2k}$,
we can use the space $\cH_{2d-1,2k}$ as the reference space.
For $k=d$,
the dimension of $\cH_{2d-1,2k}$ is $1$, {\em i.e.}, 
is smaller than that of $\cH_{2k}$.
However, for $k \,< d$,
the dimension of $\cH_{2d-1,2k}$ is 
$\cho{2d-1}{d-k}- \cho{2d-1}{d-k-1}$,
{\em i.e.}, it is greater than that of $\cH_{2k}$.
Hence, 
by replacing the reference space by the space $\cH_{2d-1,2k}$,
we can reduce the estimation error to ${\cD}_{opt}^{d-2}$
without using the reference system.
Since
${\cD}_{opt}^{d-2}
\cong \frac{\pi^2}{4(d-2)^2}
\cong \frac{\pi^2}{(2d-1)^2}=\frac{\pi^2}{n^2}$,
an estimation error of $\frac{\pi^2}{n^2}$ 
can be achieved without the reference space.
This discussion can also be applied to the case when
$n$ is an even number.

\section{Technical details}\Label{s4}
In the following, we investigate the derivations of
equations (\ref{2}) and (\ref{3}).
Since $\Tr V^j_{\hat{g}}$ and $d (I,\hat{g})$ 
depend only on the eigenvalues of $g$ $e^{i\theta/2},e^{-i\theta/2}$,
the invariant distribution $\mu$ has the form
$\mu(d\hat{g})=\frac{1}{\pi}\sin^2\frac{\theta}{2}\,d \theta
\sin \phi_1 d \phi_1 d\phi_2$
in the following parameterization:
\begin{eqnarray*}
 \hat{g}=
\left(
\begin{array}{cc}
\cos \phi_1 & \sin \phi_1 e^{i\phi_2}\\
-\sin \phi_1 e^{-i\phi_2}& \cos \phi_1
\end{array}
\right)^\dagger
\left(
\begin{array}{cc}
e^{i\theta/2}& 0\\
0 & e^{-i\theta/2}
\end{array}
\right)
\left(
\begin{array}{cc}
\cos \phi_1 & \sin \phi_1 e^{i\phi_2}\\
-\sin \phi_1 e^{-i\phi_2}& \cos \phi_1
\end{array}
\right).
\end{eqnarray*}
Hence, the relation
\begin{eqnarray}
\int_{\SU(2)} 
f(\theta) \mu(d \hat{g}) 
=
\int_0^{2\pi} \frac{f(\theta)}{\pi}
\sin^2\frac{\theta}{2} 
d \theta \Label{3-12-20}
\end{eqnarray}
holds.
Because $d (I,\hat{g})= \sin^2\frac{\theta}{2}$ and
$\Tr V^j_{\hat{g}}
= \sum_{l=1}^j e^{i(l-\frac{j+1}{2})\theta}$
and by applying (\ref{3-12-20}),
we have
\begin{align*}
& \int_{\SU(2)} d(I,\hat{g})
|\Tr V_{\hat{g}}^j|^2 \mu(d \hat{g}) 
=
\int_0^{2\pi} \frac{1}{\pi}
\sin^4\frac{\theta}{2} \left(\sum_{l=1}^j 
e^{i(l-\frac{j+1}{2})\theta}\right)
\overline{\left(\sum_{l=1}^j 
e^{i(l-\frac{j+1}{2})\theta}\right)}
d \theta \\
=&
\int_0^{2\pi} \frac{1}{4 \pi}
\left(
\frac{3}{2} - 2 \cos \theta +\frac{1}{2} \cos 2\theta
\right) 
\left(
j + 2\sum_{l=1}^j (j-l)\cos l \theta \right)
d \theta .
\end{align*}
When $j\ge 2$,
this integral is equal to
\begin{eqnarray*}
&\int_0^{2\pi} \frac{1}{4 \pi}
\left(
\frac{3}{2} - 2 \cos \theta +\frac{1}{2} \cos 2\theta
\right) 
\left(
j + 2 (j-1) \cos \theta + 2(j-2)\cos 2\theta
\right)
d \theta ,
\end{eqnarray*}
which implies (\ref{2}).
When $j=1$, we have
\begin{eqnarray*}
&\int_0^{2\pi} \frac{1}{4 \pi}
\left(
\frac{3}{2} - 2 \cos \theta +\frac{1}{2} \cos 2\theta
\right) 
\! d \theta =\frac{1}{2},
\end{eqnarray*}
which implies (\ref{2}).

With regard to (\ref{3}),
we can calculate 
\begin{align*}
&\int_{\SU(2)} d(I,\hat{g})
(\Tr V_{\hat{g}}^{2k})(\Tr V_{\hat{g}}^{2k'})^\dagger
\mu(d \hat{g}) \\
=&
\int_0^{2\pi} \frac{1}{4\pi}
\left(
\frac{3}{2} - 2 \cos \theta +\frac{1}{2} \cos 2\theta
\right)
\left(\sum_{l=1}^{2k} e^{i(l-\frac{2k+1}{2})\theta}\right)
\overline{\left(\sum_{l=1}^{2k'} e^{i(l-\frac{2k'+1}{2})\theta}\right)}
\,d \theta .
\end{align*}
In the above integral, 
the term $\left(\sum_{l=1}^{2k} e^{i(l-\frac{2k+1}{2})\theta}\right)
\overline{\left(\sum_{l=1}^{2k'} e^{i(l-\frac{2k'+1}{2})\theta}\right)}$
can be expanded into many terms;
however,
all other terms except the constant term, $2\cos \theta$, and 
$2 \cos 2\theta$ vanish.
That is, the coefficients of only the above three terms are
important.

In the following, we calculate this integral in the three cases:
(i) $k=k'$, (ii) $|k-k'|\,> 1$, and (iii) $|k-k'|=1$.
Case (i) has already been calculated in (\ref{2}).
In case (ii), these three coefficients coincide.
Thus, the integral is equal to $0$.
Next, we proceed to the case (iii).
When $k'=k+1$,
the term $\left(\sum_{l=1}^{2k} e^{i(l-\frac{2k+1}{2})\theta}\right)
\overline{\left(\sum_{l=1}^{2k'} e^{i(l-\frac{2k'+1}{2})\theta}\right)}$
can be expanded to
$k + 2 k \cos \theta + 2(k-1)\cos 2\theta+ \cdots$.
Thus, the integral is equal to $-\frac{1}{4}$.
Similarly, we can show that the integral is equal to $-\frac{1}{4}$ 
when $k'=k-1$.
Therefore, we obtain (\ref{3}).

\section{Concluding remark}
In this paper, we derived a remarkable relation between
the estimation of the $\SU(2)$ action and phase estimation.
By using this relation, we showed that the optimal estimation error 
is less than $\frac{\pi^2}{n^2}$.
The essence of this relation lies in the relation between
the two similar relations (\ref{3-12-8}) and (\ref{3}).
Indeed, 
since $\sin^2 \theta= 1- |\frac{e^{i\theta/2}+e^{-i\theta/2}}{2}|^2$,
the cost function of phase estimation is equal to 
$1- |\chi_2(\theta-\hat\theta)|^2$,
where 
$\chi_2(\theta)$ is the half of the character of the two-dimensional 
representation 
$e^{i\theta} \mapsto 
\left(
\begin{array}{cc}
e^{i\theta/2}&0\\
0&e^{-i\theta/2}
\end{array}
\right)$.
Hence, the essence of (\ref{3-12-8}) is the equation 
\begin{eqnarray}
 \int_0^{2\pi} 
(1- |\chi_2(\theta)|^2)
\chi^k(\theta)
\overline{\chi^l(\theta)}d \theta
= \frac{1}{2}\delta_{k,l}- \frac{1}{4}\delta_{k,l-1}
- \frac{1}{4}\delta_{k-1,l},\Label{3-12-12}
\end{eqnarray}
where $\chi^l(\theta)$ is the character of
the one-dimensional representation of
$e^{i\theta} \mapsto e^{il\theta}$.

On the other hand,
when we let $\chi^l(g)$ assume the character of 
the $l$-dimensional irreducible representation
of $\SU(2)$ over dimension $l$,
the essence of (\ref{3}) is
\begin{eqnarray}
 \int_{\SU(2)}
(1- |\chi^2(g)|^2)
\chi^{2k}(g)
\overline{\chi^{2l}(g)}\mu(d g)
= \frac{1}{2}\delta_{k,l}- \frac{1}{4}\delta_{k,l-1}
- \frac{1}{4}\delta_{k-1,l}.\Label{3-12-11}
\end{eqnarray}
Therefore, the main reason for the 
relation between the two estimation problems 
is the character formulas (\ref{3-12-11}) and (\ref{3-12-12}).
In other words,
the integrals 
with regard to characters are the essence of phase estimation 
as well as that of the estimation of $\SU(2)$ action.
Physically, the square-error estimators in both scenarios
can be realized by the effective use of interference between different
irreducible representations.

In this paper, when we consider the irreducible space $\cH_{n,k}$ of
the action of the permutation group as the reference space
of the irreducible space $\cH_{k}$ of $\SU(2)$ action,
we proved the existence of a square-error estimator in the estimation
of the $\SU(2)$ action without use of the real reference system.
Since the irreducible space $\cH_{n,k}$ of
the action of the permutation group has
a greater dimension than the corresponding 
irreducible space $\cH_{k}$ of the $\SU(2)$ action,
except for the exceptional cases,
the correspondence between
the irreducible space $\cH_{n,k}$ and the reference space of
the corresponding 
irreducible space $\cH_{k}$ is not unique.
Hence, it is desirable to obtain a more physically 
realizable 
correspondence\cite{Ru}.
This is an interesting future study.

Finally, we should remark on the relation between our group covariance
approach and the Cram\'{e}r-Rao approach.
In the latter approach, 
we focus on the Cram\'{e}r-Rao-type bound, {\it i.e.},
the minimum weighted sum of mean square errors
under locally unbiased conditions.
As discussed in Fujiwara and Imai\cite{FI},
we often seek an input state that minimizes this bound.
Even though this bound decrease to zero with an order of $\frac{1}{n^2}$,
we cannot conclude that
there exists a sequence of estimators
whose weighted sum of minimum mean square errors
decreases to zero with an order of $\frac{1}{n^2}$.
In the state estimation, there exists a sequence of estimators
whose weighted sum of minimum mean square errors
reduces to zero at rate $\frac{C}{n}$,
where $C$ is the Cram\'{e}r-Rao-type bound.
This is because an adaptive estimator attaining bound $C$
can be selected as follows\cite{GM,HM}.
In the state estimation, we can select a set of neighborhoods
in which the minimum weighted sum of mean square errors
can be approximated to 
the Cram\'{e}r-Rao type bounds within bounded differences.
Hence, if we choose the first estimator 
whose estimate belongs to the above neighborhood
with an exponentially small error probability,
the adaptive estimator satisfies the required condition.
However, the situation 
of the estimation of the $\SU(2)$ action
differs from that of the estimation of state.
In the estimation of the $\SU(2)$ action, 
such neighborhoods depend on the number $n$
because the state family depends on the input state and 
the required neighborhoods depend on the state family.
In other words, there is a possibility 
that the radius of the neighborhoods reduces to zero
in proportion with the number $n$ of actions.
Hence, we cannot directly obtain 
the estimation error based on 
the optimal Cram\'{e}r-Rao-type bound.

On completion of this research, the author found that the same results 
were obtained by two other groups\cite{Bagan-1,Bagan,Chiri}.
However, this approach is different from those employed by
them because it is based on the notable relation between
the $\SU(2)$ estimation and the phase estimation.
Further, Chiribella {\it et al.}\cite{CDF}
proved 
the optimality of the error 
$\frac{1}{2}\left(
1- \sum_{k=1}^{d-1} |x_k||x_{k+1}|
\right)$
while this is not proved in this paper.
They also elucidated out that
character integrals and interference of irreducible representations 
are very useful for finding 
the optimal estimation of an unknown group transformation not
only for $U(1)$ and $\SU(2)$ but also for arbitrary groups.

Further, on completion of this paper,
the author also found Acin {\it et al}.'s paper \cite{AJV}.
They almost mentioned that 
the optimal initial state has the form 
$x_{\vec{x}_d}^{2d-1}$.
However, they did not derive the equation corresponding to (\ref{3-15}).
In addition, Rudolph and Grover \cite{RG}
discussed a similar problem from a computational viewpoint.

\section*{Acknowledgments}
The author wishes to thank Professor Keiji Matsumoto,
Dr. Yoshiyuki Tsuda, and Mr. Aram W. Harrow for their useful discussions.
The author is also grateful to Professor Hiroshi Imai and
all the members of ERATO Quantum Computation and Information Project
for their kind support.
The author would also like to 
thank Dr. Terry Rudolph for kind comments for concluding remark.
The author is indebted to the reviewer for some useful comments.

\end{document}